\documentclass[12pt,preprint]{aastex}

\newcommand{\Msun}{\rm M$_\odot$}

\shorttitle{Inner structure of the TW Hya Disk in scattered light} 
\shortauthors{Debes et al.}
\begin{document}
\title{The inner structure of the TW Hya Disk as revealed in scattered light\footnote{Based on observations made with the NASA/ESA Hubble Space Telescope, obtained from the data archive at the Space Telescope Science Institute. STScI is operated by the Association of Universities for Research in Astronomy, Inc. under NASA contract NAS 5-26555.}}
\author{John H. Debes\altaffilmark{1}, Hannah Jang-Condell\altaffilmark{2}, Glenn Schneider\altaffilmark{3}}
\altaffiltext{1}{Space Telescope Science institute, Baltimore, MD 21218, U.S.A.}
\altaffiltext{2}{Department of Physics \& Astronomy,  University of Wyoming, Laramie, WY 82071, U.S.A.}
\altaffiltext{3}{Department of Astronomy, The University of Arizona, Tucscon, AZ 85721 ,U.S.A.}

\begin{abstract}
 We observe a significant change in the TW Hya disk interior to 40~AU via archival unpolarized multi-wavelength Hubble Space Telescope/STIS and NICMOS images with an inner working angle (IWA) of 0\farcs4 (22~AU).  Our images show the outer edge of a clearing at every wavelength with similar behavior, demonstrating that the feature is structural, rather than due to some property of polarized light in the disk.  We compare our observations to those taken by \citet{a15} and \citet{rapson15}, and discuss the spectral evolution of the disk interior to 80~AU.  We construct a model with two gaps: one at 30~AU and one at 80~AU that fit the observed surface brightness profile but overpredicts the absolute brightness of the disk.  Our models require an additional dimming to be consistent with observations, which we tentatively ascribe to shadowing.  The gap structures seen in scattered light are spatially coincident with sub-mm detections of CO and N$_2$H$^+$, and are near expected condensation fronts of these molecular species, providing tentative evidence that the structures seen in scattered light may be correlated with chemical changes in the disk.
 \end{abstract}

\keywords{Stars: Circumstellar Matter, Stars: Planetary Systems: Formation, Stars: Planetary Systems: Protoplanetary Disks, Stars: Individual: TW Hydrae}

\section{Introduction}
TW Hya is the closest protoplanetary disk system (d=54~pc) and extensively studied \citep{vanLeeuwen:2007}.  Its disk is nearly face-on (i$<$7$^\circ$; \citet{qi04}) and shows hints of active planet formation despite its 10~Myr age \citep{bergin13}.  Estimates of its stellar mass range from 0.5-0.8 M$_{\odot}$ \citep{alencar02,Vacca,herczeg14}, making the TW Hya disk an analog for the Solar Nebula.  Recently, a CO snow line at 30~AU has been discovered with ALMA maps of N$_2$H$^+$ \citep{qi13}.   Concurrently, far-IR spectra of water vapor at 4~AU are indicative of a snow line moving outwards as the disk ages \citep{zhang13}.  Hubble Space Telescope (HST) has observed the disk with increasing fidelity, including WFPC2 imaging (IWA$\approx$0.7\arcsec; Krist et al., 2000), NICMOS F110W and F160W imaging \citep[IWA=0\farcs38][]{Weinberger:2002}, and STIS 50CCD imaging and spectroscopy \citep[IWA$\approx$0.6\arcsec][]{Roberge:2005}.  We recently used seven HST images of the disk spanning 0.5-2.2\micron\ to infer the presence of a partially cleared gap at 80~AU \citep[][hereafter D13]{debes13}.   That study reported surface brightness (SB) profiles to $\sim$40~AU as a function of wavelength.  In this letter we observe the disk inward to 22~AU, the IWA of the NICMOS coronagraph.  In \S \ref{sec:conf} we recover the inner structure of the SB first noted in \citet{Weinberger:2002} and further reported in \citet[hereafter A15]{a15} and \citet[][hereafter R15]{rapson15} in seven filters of visible and near-IR light.  In \S \ref{sec:model} we present self-consistent models of the inner region of the disk in scattered light that demonstrate the change in SB is not due simply to a gap structure, but a more complex interplay between a depression in SB around 30~AU and unseen inner structures that are also depressing the expected brightness of the disk.  Finally, in \S \ref{sec:disc} we discuss the implications of our findings in the context of other recent protoplanetary disk studies.

\section{Confirmation of surface brightness changes interior to 40~AU}
\label{sec:conf}
D13 details the process of re-reducing the NICMOS and STIS datasets, a discussion of the outer gap detected at 80~AU, and the impact of dust composition on the observed SB profiles.  D13 chose only to extend their analysis of the disk SB profiles to within 0\farcs7 to include information gained from STIS-derived spatially resolved spectra of the disk.  Our re-analyzed observations include seven wide- and medium-band filters that span wavelengths from 0.5~$\micron$-2.22$\micron$, which were azimuthally averaged and extend to an IWA=0\farcs4 (22~AU).  Photometric uncertainties were determined from PSF-subtraction, background, read-noise, and dark rate, and the standard deviation of counts within the annulus used to calculate the azimuthal average.

\subsection{Surface Brightness Profiles}
\begin{figure}
\plottwo{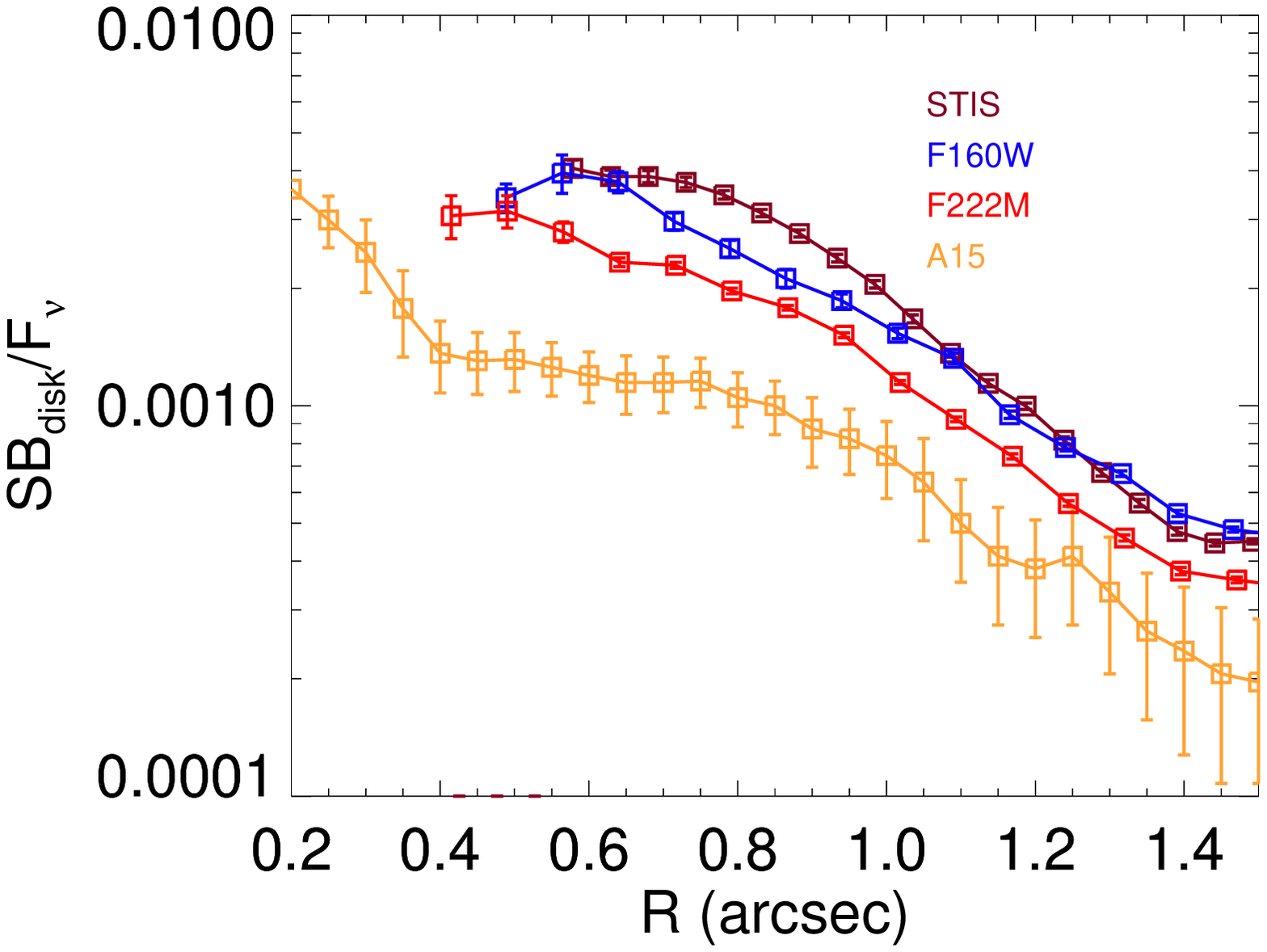}{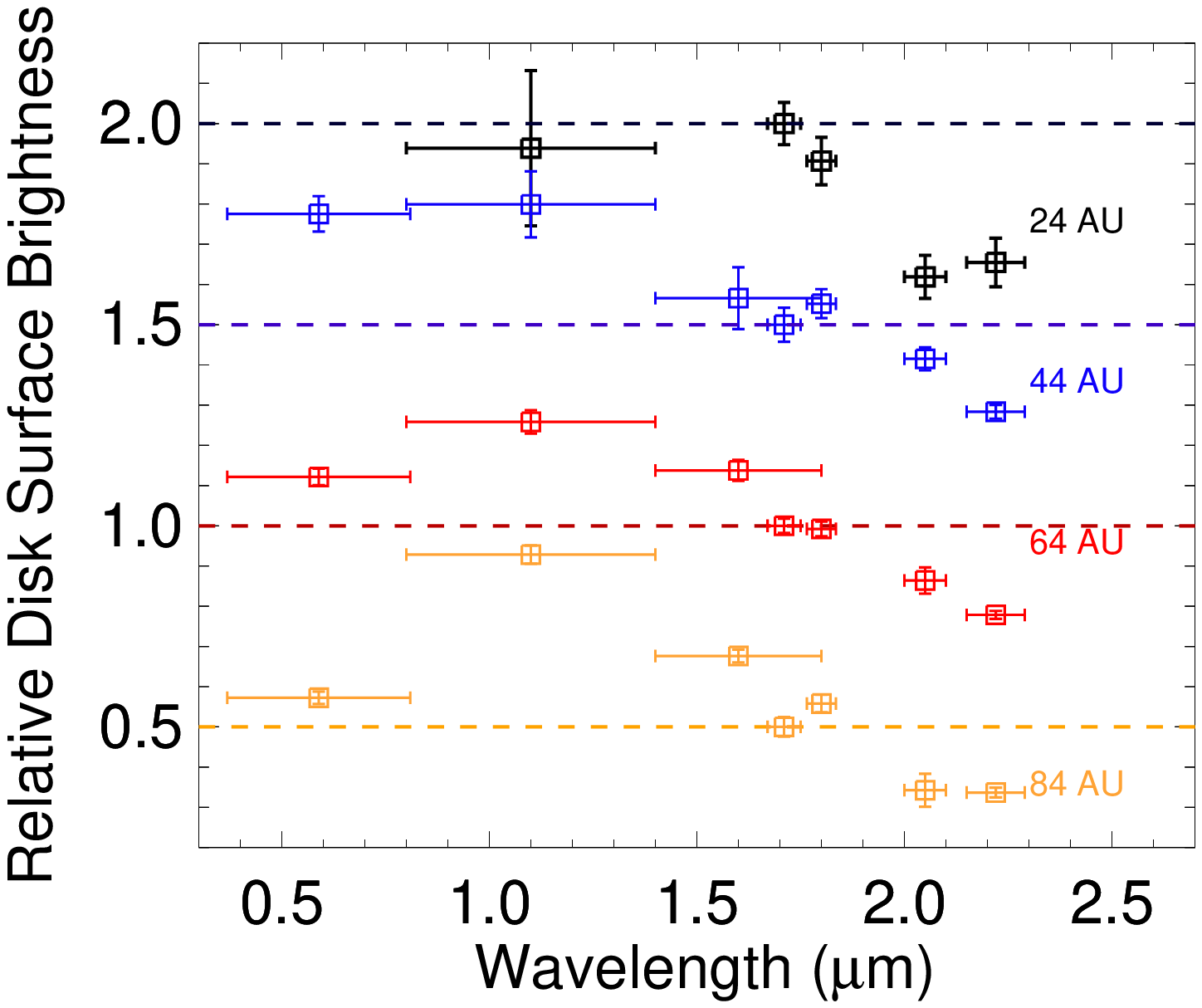}
\caption{\label{fig:sbprof}(left) SB profiles of the TW Hya disk.  We show profiles from three HST images of TW~Hya (STIS, F160W, F222M) and the profile from A15.   The polarized intensity observed by A15 is 77\% of the total intensity observed in F160W.  (right) Photometry of the disk as a function of wavelength and distance from the central star.  At each distance the normalized spectrum is offset by a constant factor.  A depression in flux in 2.04 and 2.22\micron\ strengthens interior to 80~AU.}
\end{figure}

Here we investigate radii $<$1\farcs5 in the HST data to determine the significance of the gap reported by A15 and R15.  To confirm the observed changes we revisit the SB profiles for TW~Hya for each spectral band, presented in Figure \ref{fig:sbprof}.  A15 report a power-law drop in polarized light with r$\propto-1.39$ from their inner working angle to 0\farcs4 (Zone 1), followed by a shallow power-law slope to the SB from 0\farcs4 to 0\farcs8 with a value of -0.33 (Zone 2).  Exterior to 0\farcs8 out to 1\farcs5, the SB drops again more steeply (Zone 3; -2.65).  As can be seen in Figure \ref{fig:sbprof}, we replicate the same shallow behavior to the SB in unpolarized light across each of our filters in Zone 2, including wavelengths that overlap with the $H$ filter bandpass.   

We calculated power-law slopes for each spectral band.  The median power-law slope of Zone 2 in all bands is -0.9$\pm$0.2 (1-$\sigma$), which is marginally consistent to A15's slope given our uncertainties.  For Zone 3, we find slopes in the NIR range from -3.7 to -2.9, with no obvious trends associated with wavelength.  As the individual uncertainties in the slopes roughly match the standard deviation of the reported values, this suggests that the scattered light of the inner disk is not changing significantly with wavelength in the NIR.  The median NIR power-law slope for Zone 3 is -3.0$\pm$0.2, which is marginally steeper than that found by A15 and less steep than that reported in R15.  The measured slope in the STIS band is much steeper, at -3.7$\pm$0.2.  Based on our current inner working angle, we are unable to observe the rise in SB reported by A15 inwards of 0\farcs4, but that behavior has been confirmed in R15.

\subsection{Spectrum of the inner disk}
 A15 postulated that the structure observed may be due to grain growth, an increasing temperature in the inner disk, or through changes in surface density.  D13 also investigated some of these possibilities.  D13 found that the disk color was unchanged with stellocentric distance, but noted a tentative difference in color within the gap region around 80~AU.  Here, we re-calculate the 0.5-2.2\micron\ spectrum at four radii within the disk and look for any trends.   

We calculated the spectral intensity relative to the star's brightness by dividing each SB profile by the calculated photometry of TW Hya as detailed in D13.  We then constructed a two-dimensional spectrum in wavelength vs. radius and normalized the SB at each radius to the value in the F171M images, since the disk was well detected at all radii in this band.  We averaged the normalized SB between 0\farcs4-0\farcs44 ($\sim$24~AU), 0\farcs66-0\farcs96 ($\sim$44~AU), 1\arcsec-1\farcs34 ($\sim$64~AU), and 1\farcs4 to 1\farcs7 ($\sim$84~AU), which encompasses the inner flattening of the SB, the transition to the 80~AU gap location, and the gap itself.  Spectra of the outer disk are presented in D13, as well as discussion of how composition and grain size can change the observed SB profile of a disk.  

We note that the F204M and F222M photometry show a trend of being more depressed toward both interior and exterior to the 80~AU gap, with an inflection point at the approximate gap location.  In the interior, the depression in both bands reaches 60\% the brightness in the F171M bandpass, hits a maximum of 80\%, and then decreases steadily to the outer disk.  Our D13 models also predict the increasing depression of these filters for small dust, primarily due to an overall decreasing scattering efficiency with increasing wavelength for small grains as well as due to mild water ice absorption at 2.0 \micron.  The deepening of the depression interior to the gap is not predicted by our models and is accompanied by a flattening of the spectrum $<$1.6\micron.  A full explanation is beyond the scope of this paper, but these two behaviors are qualitatively seen for the D13 best fit model composition for TW Hya as grain size increases.  These trends are mildly correlated with the observed SB profiles, but do not show any sharp transitions, suggesting that dust grain property changes are not the sole cause of the SB behavior.  Higher resolution near-IR spectra of TW Hya's disk could potentially identify water ice or methane ice in the disk surface layers.  We next turn to explaining the inner SB profile with an additional gap, similar to the one proposed at 80~AU in D13.
 
\section{Modeling of a gap in the interior of the TW~Hya disk}
\label{sec:model}
To model the TW Hya disk, we use the same parameters and composition as used in D13.  
To wit, $\alpha=0.005$,
$\dot{M}=10^{-9}\,M_{\odot}\mbox{ yr}^{-1}$, 
$M_*=0.55\,M_{\odot}$,
$T_{\rm eff}=3741$ K, and 
$R_*=1.08\,R_{\odot}$.
The disk model is constructed as described in D13,
but over the radial range $10-70$ AU, 
interior to the gap at 80 AU modeled in D13.

The gap is modeled as a Gaussian-shaped perturbation to the surface density profile of the disk, as
\begin{equation}
\Sigma(r) = \Sigma_0(r) \{1-d\exp[-(r- r_{0})^2/(2w^2)] \}
\exp[ -(r/k) ]
\end{equation}
with depth $d = 0.3$, width $w= 10$ AU, and centered at $r_0=30$ AU\@. 
We investigated how this gap would affect the brightness
profile, both alone and in conjunction with a second outer gap at 80 AU
as modeled in D13.  

We model the temperature and density in both the radial and vertical
dimensions, assuming axisymmetry.  This is done with a radiative transfer code under the assumption of
hydrostatic equilibrium, taking disk self-shadowing in the gap into account \citep{jangcondell12,jangcondell13}
The temperature and density are
calculated iteratively to account for the feedback of radiative
heating and cooling on the disk structure resulting from shadowing and
illumination of the gap in the disk.  

D13 showed that differences in stellar mass on the order of the mass uncertainties for TW~Hya result in disk brightnesses that change by $\sim$5-10\%, with higher masses dimming the disk.  The latest mass estimates from \citet{herczeg14} place the most likely mass of TW~Hya at 0.69\Msun, compared to 0.55$\pm$15\Msun\ in D13.  However we wish to present results that can be directly compared to D13's results and thus the absolute brightness of the disk is slightly uncertain at this level, though not the shape of the SB distribution nor the relative brightnesses between different model density distributions.  

The disk brightness is calculated at a wavelength of 1.6~\micron.  The disk scattered light profiles shapes vary slightly with wavelength, but differences between 0.5-2.22\micron\ are minor.  Since we are interested in the broad shape of the SB, we save more detailed modeling for a future paper.

The gap-free model for the TW Hya disk has a broken
SB profile, with the break at around 13 AU\@ (See Figure 2, right panel).
This is because the primary heating sources in the disk model are
viscous heating at the midplane and stellar irradiation.  The viscous
heating creates a steeper temperature profile than radiative heating,
so the SB profile is correspondingly steeper at this radius.
Beyond around 13 AU, radiative heating dominates, so the surface
brightness profile is shallower.  For this reason, all our profiles show
a steeper brightness profile in the innermost part of TW Hya's disk.  This aspect of protoplanetary disks should be observable for other stars as coronagraphs achieve smaller IWAs.

We find that a 
simple gap centered at 30 AU (Figure 2, green line) successfully 
reproduces the shape of the brightness profile in the 30-80 AU region, 
however, the brightness is overall three times higher than TW Hya.  When the 30 AU gap is imposed on the best-fitting disk 
model found in D13, that is, a gap at 80 AU and truncated at 100 AU 
(Figure 2, blue line), the brightness profile also matches, but is a factor of 2 too bright.  Preliminary results indicate that a 
deeper inner gap will raise the brightness of the irradiated outer 
edge of the gap, while wider gaps will shift the radial location of 
the irradiated edge outwards.  

None of the models reproduce the brightness deficit interior to
20 AU observed in A15 and R15.  This is a limitation of our
radiative transfer disk models, which assume a continuous disk
interior to the simulation boundary condition.  The depth
of the brightness deficit suggests strong shadowing, possibly
from the inner wall at 4 AU\@, a conclusion also reached by R15.  In order to 
adequately model the TW Hya disk scattered light SB profile's absolute brightness, we need to include all 
aspects of the disk structure from 4 AU to 100 AU\@.

\section{Discussion}
\label{sec:disc}

We present a re-analysis of the interior regions of TW~Hya with high contrast imaging and confirm the presence of a gap-like feature in the disk interior to 40~AU, with a probable center at close to 30~AU.  Based on initial qualitative models, we show that the gap is not fully cleared, and that there may be additional structure or shadowing interior to 20~AU.  TW Hya, just like other protoplanetary disks with inner cavities, reveals a complex structure with radius that is qualitatively different for larger dust, sub-micron sized dust, and the gas in the disk.

TW Hya's SB distribution can be contrasted with other disks.  Some disks show clearings at long wavelengths but no evidence for a clearing at shorter polarized wavelengths such as SR 21 \citep{follette13}.  Some disks show different gap sizes between wavelengths, such as SAO 206462 \citep{muto12,garufi13}.  Other disks show large clearings ($>$10~AU) both in their SED and from scattered light, like PDS 70 \citep{dong12a}.  Others show clear gap-like features similar to TW~Hya, such as HD~169142 \citep{quanz13,momose15}.

We modeled the TW~Hya disk with the methods laid out in \citet{jangcondell13}--with gaps that have a shape similar to those produced by forming proto-planets, but the interior SB profile of the disk departs from the predictions of \citet{jangcondell13} for single planets.  Further, our models predict very modest gaps or features in ALMA images, since we are inferring partially cleared gaps from the observed scattered light SB profiles.  TW~Hya's disk structure is now reminiscent in scattered light to the sub-mm emission reported for HL~Tau \citep{hltau}, which shows at least three convincing gap structures at the disk midplane.  For TW~Hya the picture is more complex because our data samples the upper layers of the disk \citep[D13][]{juhascz15}, and could also be created by dust evolution \citep{birnstiel15}.  The location of HL Tau's gaps may correlate to condensation fronts in the HL~Tau disk.  This may signal where pebble accretion is occurring most efficiently due to the presence of pure water ice or clathrate hydrates which trap volatile species in crystalline lattice networks \citep{blum08,wada09,gundlach11,johansen,testi,zhang15}.    

For TW~Hya, its gaps correlate with condensation regions of CO and N$_2$ ice.  The midplane temperatures around the regions of both gaps correlate with the condensation temperature of these volatile ices.  For comparison, we show the pure CO condensation temperature range which overlaps with the radial location of the inner gap in Figure \ref{fig:modelt} \citep{zhang15}.  N$_2$H$^+$ emission is thought to be a tracer for CO condensation below temperatures of about 16-20~K, which corresponds to radii of 25-30~AU in our model and comparable to the inner edge of N$_2$H$^+$ gas detected by \citep{qi13}.  Further, this gas will dissipate as N$_2$ condensation occurs, which occurs at below 12-15~K \citep{zhang15}.  The outer edge of the observed N$_2$H$^+$ ring is 100~AU, where the SB of TW~Hya recovers from the 80~AU gap.  Therefore, the outer edge of the 80~AU gap could be related to a region of the disk where N$_2$ condensation is occurring.  

Based on these features, we would predict a shallow inner gap at 30~AU in ALMA observations of the TW~Hya disk if the gap were caused by condensation fronts or a protoplanet.  Spectral indices for dust in the gaps may be similar to those reported for HL~Tau, where the inferred spectral index of the dust (I$_\nu \propto \nu^\alpha$; $\alpha\sim2$) is consistent with very large grain growth in the interior, which is hinted at by our spectral observations of TW Hya \citep{hltau}.  HL~Tau shows a further prominent gap at the H$_2$O condensation front.  Based on the detection of a water ice line at 4~AU \citep{zhang13}, the inner clearing of TW~Hya may be related to that distance and its structure may also be detectable if directly imaged in scattered light.  ALMA observations of CO gas relative to dust continuum in the interior could also potentially determine whether both gaps were due to planets, or whether the outer gap was due to dust fragmentation and pile-up, such as has been recently seen in dust and gas coupled simulations of disks \citep{gonzalez15}.

VLA observations of TW Hya have previously revealed the presence of a 4 AU
inner hole (Hughes, et al. 2007).  SMA observations confirm the presence
of this inner hole as well as a truncation of the dust disk at around
60 AU (Andrews, et al. 2012), roughly consistent with the 80 AU gap
seen in scattered light (D13).  However, the gap at 30 AU modeled here,
with a depth of 0.3 and width of 10 AU, is too shallow and narrow to
be recovered by either observation.  At a distance of 54 parsecs,
10 AU subtends only 0\farcs18, whereas the beam size of the SMA used to
observe TW Hya is 0.80\arcsec$\times$0.58\arcsec.  The VLA observation used a smaller
beam size, but had less sensitivity to emission at distances of 30-80~AU.  As a test, we
generated synthetic images of the thermal dust emission
from our models of TW Hya using the
opacities in D13 and the radiative transfer methods from
\citep{jangcondell12} and convolved the images 
with a Gaussian beam matching that of the SMA\@.
We found that the addition of a gap described above
yields at most a 5\% decrease in the brightness profile.  This is reasonably consistent with recent results of \citet{hoger15}, who also suggest that gaps with depths $<$ 80\% deep are still allowed with their ALMA data of TW~Hya.  Concurrently, a tentative observation of a gap in TW~Hya's dust continuum with ALMA has
recently been reported by \citet{nomura15}, which if confirmed, could be consistent with our scattered light images and sub-mm model predictions. 
Regardless, differences between the scattered light brightness of a
disk and the dust emission from sub-millimeter observations
can be affected by processes such as
grain growth, dust settling, and disk turbulence,
just to name a few.  A hydrostatic well-mixed disk model, as used
in this work, cannot capture these effects.

The SB profile observed could be produced by an inner rim \citep{dong15}.  An inner rim creates a smoothly transitioning SB profile due to self-shadowing and we require some shadowing to explain our models' overprediction of flux.  Some disks even show rapid variability \citep[e.g][]{wisniewski08}, which could be caused by shadowing.  If the shadowing is only influencing a limited range of radii, one might expect variability in the scattered light over only those radii, and that the SB profile might change with time.  Additional observations of TW~Hya in the same wavelength across longer baselines is needed to further test this idea.
  
High-contrast scattered-light imaging at optical and nIR wavelengths at the small IWAs enabled with HST and ground based facilities such as Subaru, GPI, and SPHERE, will reveal new mysteries about the inner TW~Hya system--with the potential promise of understanding planet formation at radii interior to 20~AU and linking up with interesting dust structures observed with ALMA.

\acknowledgements
The authors would like to thank Alycia Weinberger for kindly providing her reduction of the STIS images of TW Hya and helpful discussions related to its disk.  We also wish to thank Thayne Currie for providing the surface brightness profile for TW~Hya as obtained in A15.


\begin{figure}
\plottwo{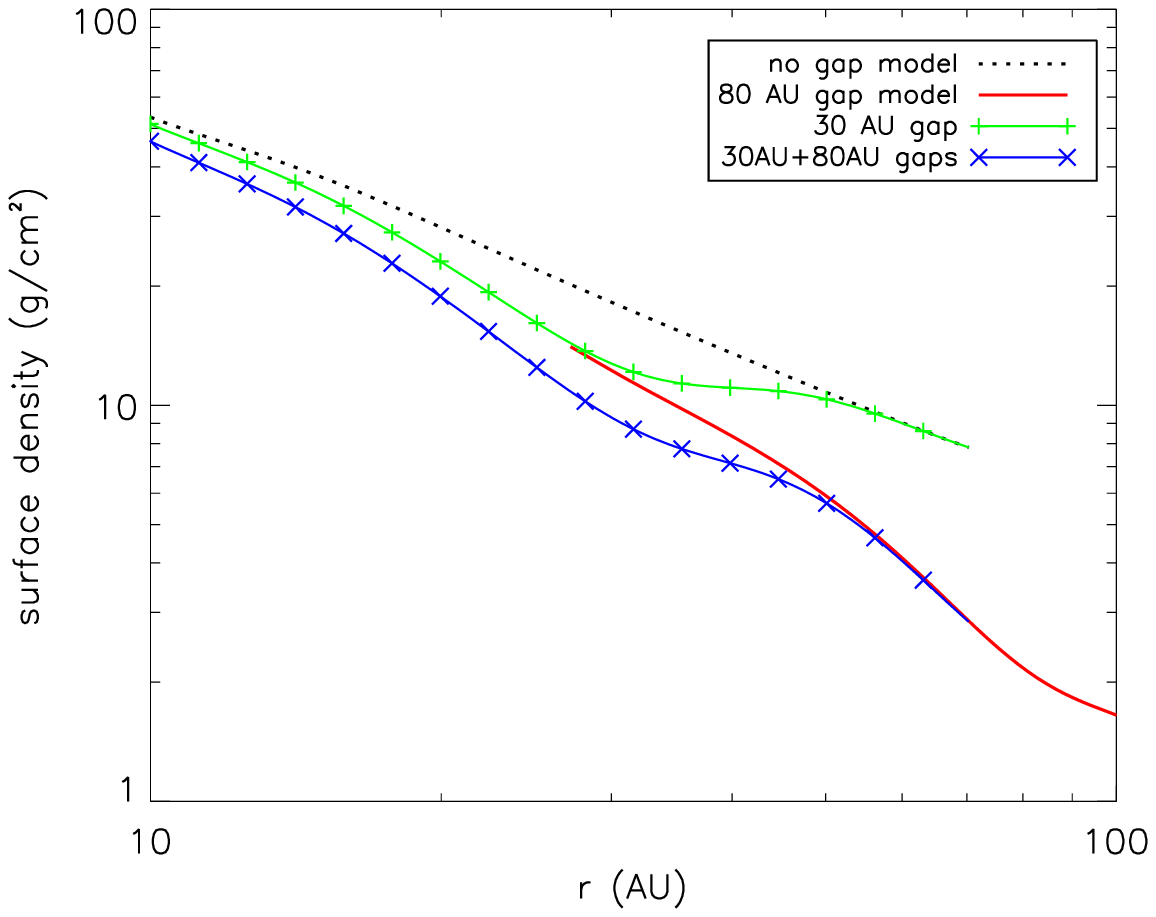}{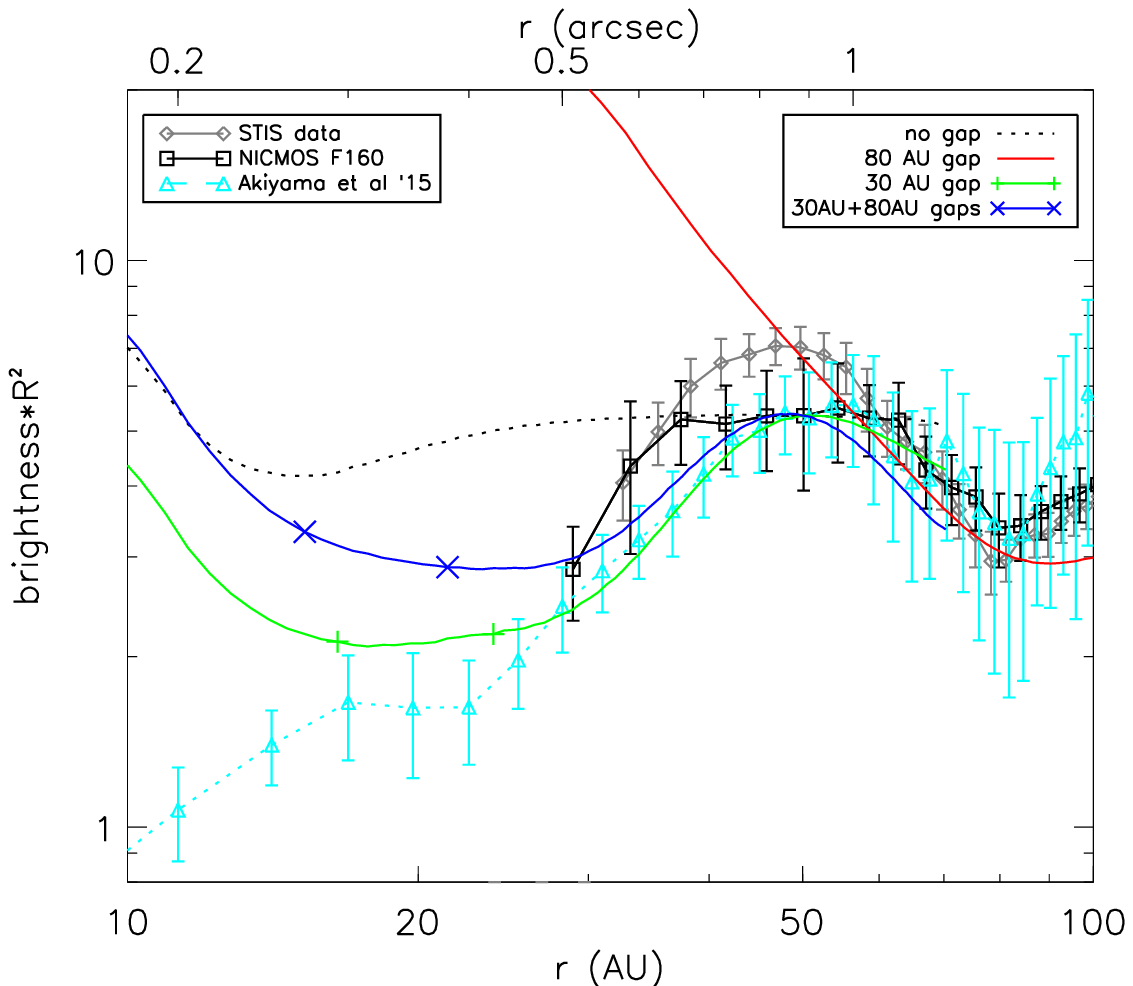}
\caption{\label{fig:models} (left) Modeled surface density distributions of the TW~Hya disk from D13 (red line), no gap (dashed line), and our two interior gap models.  (right) Modeled SB distributions of the TW~Hya disk for the same models in the left panel multiplied by $r^2$ to enhance gap features.  For this figure we have scaled our single 30 AU gap model by a factor of 0.28 and our two gap model by a factor of 0.50 to match the F160W surface brightness profile.  We also scaled the resulting the A15 surface brightness profiles by 1.3 to match the F160W surface brightness profile.  A disk with an inner 30~AU gap and an outer 80~AU gap qualitatively matches the observed behavior from 30-100~AU but overpredicts disk flux. The marked points on the single and two gap models represent the boundaries of the CO condensation temperature range shown in Figure \ref{fig:modelt}}
\end{figure}

\begin{figure}
\plotone{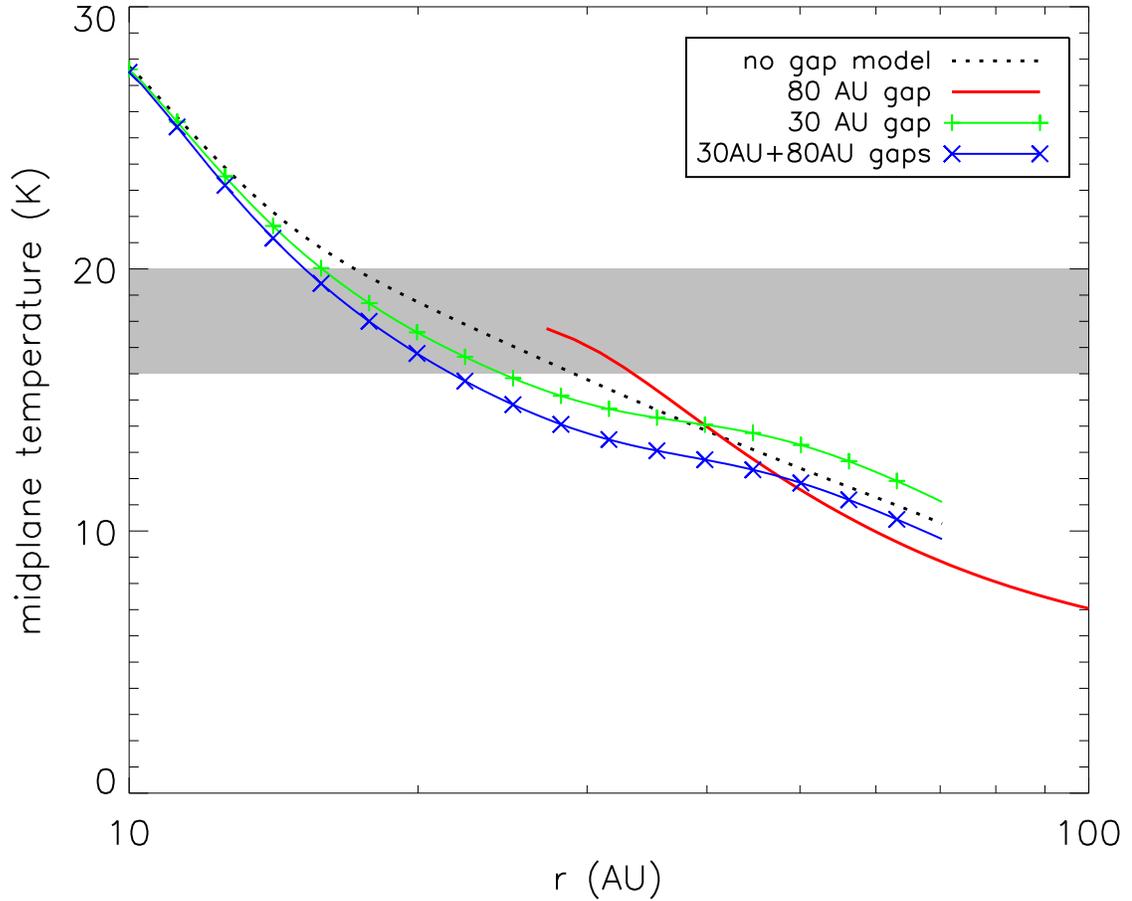}
\caption{\label{fig:modelt}Midplane temperatures of the TW~Hya disk under the assumption of various disk structures.  Midplane temperatures are only weakly dependent on surface structures such as partially cleared gaps.  We also overplot the expected CO condensation temperatures, which correlate roughly with where we observe the inner disk gap.}
\end{figure}

\end{document}